\documentclass{elsart}
\usepackage{graphicx}

\begin{document}

\begin{frontmatter}
\title{SIS epidemics with household structure: the self-consistent field method}  
\author{G. Ghoshal, }
\author{\corauthref{cor1}L. M. Sander,}
\corauth[cor1]{Corresponding author. email lsander@umich.edu}
\address{Michigan Center for Theoretical Physics, Department of
Physics, University of Michigan, Ann Arbor, Michigan - 48109, U.S.A.}
\author{I. M. Sokolov} \address{Institut f\"ur Physik,
Humboldt-Universit\"at zu Berlin,  Newtonstr. 15, 12489
Berlin, Germany}

\begin{abstract} 
We consider a stochastic SIS infection model for a population partitioned into
$m$ households assuming random mixing. We solve the 
model in the limit $m \to \infty$ by
using the self-consistent field method of statistical physics. We 
derive a number of explicit results, and give numerical illustrations.
We then do numerical simulations of the model for finite $m$ and 
without random mixing. We find in many of these cases that the self-consistent
field method is a very good approximation. 
\end{abstract}
\begin{keyword}
SIS epidemic \sep Households \sep Stochastic models \sep Statistical
physics
\end{keyword}

\end{frontmatter}

\section{Introduction}
In this paper we will give a treatment of an SIS infection model in
a population partioned into households. Our main emphasis will be
the use of a method borrowed from statistical physics, the self-consistent field
(SCF) technique.  

The susceptible-infected-susceptible (SIS) model is one of the 
basic themes in mathematical epidemiology \cite{Ross,Hethcote}. 
In this idealization
of the spread and persistance of an infection, $N$ individuals are initially
either 
susceptible, $S$, or  infected, $I$. Each one of the infected 
is coupled to a certain
number of other agents. Each $I$ infects all of the $S$ with which
it has contact with a rate $B$, and each $I$ recovers with rate
$\gamma$, and immediately becomes susceptible again. 
Schematically, we can write:
\begin{equation}
S + I  \to 2I \qquad 
I \to S  \nonumber
\end{equation}
Clearly, at long times the system might achieve an equilibrium
where the number of infected fluctuates around a mean. We characterize this
state by denoting  the mean fraction of population which is infected
by $f$, i.e., 
\begin{equation}
f=<I>_t/N, 
\end{equation}
where $<>_t$ denotes a
time average. We refer to this quantity as the endemic level. 
 
The model is defined by the two parameters, $B, \gamma$, and by the contact
structure of the population. Two cases of the latter have been studied
in some detail. In  the first, each individual is equally coupled to all the
others in a large population.
 This random mixing model is rather easy
to treat, and a large number of results are known \cite{Hethcote}.
For example, in the deterministic limit, $I$ obeys: 
\begin{eqnarray}
 dI/dt &=& bIS/N - \gamma I \nonumber \\
 S &=& N -I
\label{deter}
\end{eqnarray}
where we have redefined $B=b/N$. It is easy to show that for 
$b>b_c \equiv\gamma$, and
for any intial condition with  $I>0$, the epidemic goes to 
the endemic level,
$f=1-\gamma/b$. For $b<b_c$ the infection dies out.

In the second well-studied case the
individuals are arranged in some geometric way and couple only to their
nearest neighbors. This is called the contact process \cite{contact}.
In the last section of this paper we give numerical results on a 
model which interpolates between the two cases.
 
Ball \cite{Ball} introduced the SIS household model, which puts more
structure into the random mixing case by having two levels of mixing.
In this 
idealization the population, $N$, is partitioned into $m$ households
with $n$ members. The infection rate within the household, $w$, is different
from that with individuals outside, $B$. 
This followed work on SIR models with similar structure \cite{sirh}.
Ball's results are for the case $m \to \infty$ with fixed $n$, that is,
a large population of small households, which is clearly of practical 
interest.

In this paper we extend Ball's work from a different point of view inspired
by techniques in statistical physics. These will be explained in the next
section. Briefly, the idea is  that for random mixing and a large population,
we can consider that each household is acted on by an average force of
infection from outside. This is quite similar to methods used in the 
theory of magnetism \cite{Huang}, inspired by the work of P. Weiss. 
Weiss introduced the `molecular field', which is proportional to
the mean orientation of  the neighbors
of a given magnetic ion. The orientation of the ion itself is then 
determined, and the unknown molecular field is gotten by demanding that
the neighbors have the same orientation. (Note that this
step assumes that there is no correlation between neighbors.) 
This step is known as self-consistency, hence the name, self-consistent field. 
As we will see, $f$ plays the role of the self-consistent field.  The technique
itself is called  self-consistent field (or sometimes mean-field) theory.

Some of our results were already obtained in \cite{Ball}, e.g., the 
expression for $b_c$ for households of size $n$, Eq. (\ref{thresh}), below. 
Other explicit results are new. We think that one of the main values of
this paper is that this point of view allows us to look at the 
solutions in a new way, and gives considerable insight.

In the last section we continue the analogy to magnetic systems by
doing numerical simulations for finite $m$, and for systems where the
random-mixing assumption is not valid. We find, as in magnetism, that
the SCF assumption is remarkably accurate in situations where we would
expect it to fail.

\section{Stochastic formulation and steady-state solutions} 
Consider an SIS epidemic in a population partitioned into $m$
households each consisting of $n$ individuals. We suppose that within
a household the infection rate per contact is $w$ and between
different households $B$. We write
$B=b/[(m-1)n]$. (In the notation of Ball \cite{Ball}, $\lambda_w/n=w,
\lambda_b=b$ ). We suppose that $b$ remains finite in the limit $m \to
\infty$. The recovery rate is $\gamma$, and a recovered
individual can be reinfected at once.

\subsection{Self-consistent field equations}
We begin intuitively by considering the case $n = 2$
and try to formulate  the Chapman-Kolmogorov forward
equations for the probability, $P{_k}$, to have $k$ infected in
a typical household:
\begin{eqnarray}
\frac{dP_{0}}{dt} &=& \gamma P_{1} - 2bfP_0 \nonumber \\
\frac{dP_{1}}{dt} &=& 2bfP^j_0 - (\gamma + w + bf)P_{1} + 2\gamma
P^j_{2} \nonumber \\
 \frac{dP_{2}}{dt} &=& (w + bf)P^j_{1} - 2\gamma P_{2} \label{P} \\
 f &=& (P_1 +2P_2)/2 \label{f}
\end{eqnarray}
The terms in $b$ are  the mean
force of infection between households, 
and $f$ has the interpretation of the fraction of
a typical household that is infected. The point is that the only
communication between households is by infection, and for 
a large number of such contacts, only the average number of infected
in the other households is important. In effect, we have assumed that 
the households are independent.
Note that $\sum_kP_k=1$, and
$0\le f \le1$. 

The strategy for solving these equations (in the steady
state) is to find the $P_k$ with $f$ as a parameter, and solve for
$f$ self-consistently, using Eq.  (\ref{f}), at the end of the
calculation. This is a heuristically appealing method.  For magnetic
models, the approximation is exact in the case of infinite-range
interactions (in our case, all households interact equally with all
others). Even for short-range interactions (i.e., households
interacting with a `neighborhood') the approximation is often
surprisingly useful.

In fact, a rigorous proof of the validity of Eqs. (\ref{P}, \ref{f})
is available for this case as well. This is given by Ball \cite{Ball}
who uses the work of Kurtz \cite{Kurtz} to prove the following: let
$X^{(m)}_k$ be the number of households in the population with exactly
$k$ infected. Define $P_k =\lim_{m\to \infty} X^{(m)}_k/m$. Then $P_k$
satisfies Eqs. (\ref{P}, \ref{f}) for any initial condition with
finite density, i.e., with $P_k(0), k>0 $ not all equal to $0$. In the
last section we will investigate numerically how good this
approximation is for finite $m$ and for finite range of interaction.

We note, for later use, that we can give another interpretation to
Eq. (\ref {P}). Suppose we consider a single household with 2
individuals subject to a \emph{fixed} external force of
infection. Then we can think of $P_k$ as the probability to have $k$
infected. From standard results on stochastic processes, there is a
unique steady-state solution to these equations for any positive $f,
b, w$. Further, if $f=0$ the only steady-state solution is $P_0=1,
P_1=P_2=0$; that is, the infection becomes extinct with probability
unity.

In the general case the time evolution of the vector of probabilities
$\mathbf{P} = \left( P_{0},P_{1},...,P_{n}\right)^T $
($^T$ denotes a column vector) to have
exactly $k$ infected in the household of $n$, is given by:
\begin{equation}
\frac{d\mathbf{P}}{dt} =\mathbf{M}(f)\mathbf{P} \label{Timedep}
\end{equation}
where $\mathbf{M}(f)$ is a square tridiagonal matrix with nonzero
elements
\begin{eqnarray}
m_{k,k-1} &=&\left( n-k+1\right) \left[ \left( k-1\right) w+fb\right]
\label{elements} \\
m_{k,k\quad } &=&-\left[ k\gamma +(n-k)kw+(n-k)fb\right] \nonumber \\
m_{k,k+1} &=&\left( k+1\right) \gamma \nonumber
\end{eqnarray}
($0\leq k\leq n$). The equations again contain $f$, the mean level of
infection, as a parameter.
\begin{equation}
f=\frac{1}{n}\sum_{k=0}^{n}kP_{k} \label{fn}
\end{equation}
For the reasons given above, these equations are valid in the limit $m
\to \infty$.

We will be interested in the stationary solutions of
Eqs. (\ref{Timedep}, \ref{fn}). It is clear by inspection that there
is always a trivial steady-state solution $f=0, P_0=1, P_k=0, k>0$
corresponding to extinction of the infection. We will find that above
a certain threshold there is another solution, and
we will show that when this occurs the trivial solution is unstable.

\subsection{Steady-state solutions}

In the case $n=2$ it is a matter of simple algebra to find the
steady-state solution:
\begin{eqnarray}
P_{1} &=& \frac{2fb\gamma}{\gamma^2 + 2fb\gamma +(fb)^2 + fbw}
\nonumber \\ P_{2} &=& \frac{(fb)^2 + fbw}{\gamma^2 + 2fb\gamma
+(fb)^2 + fbw}
\end{eqnarray}

Using Eq. (\ref{f}) we get a cubic polynomial equation to solve for
the endemic level, $f$. After removing the trivial solution we have:
\begin{equation}
(fb)^2 + f(2b\gamma + wb - b^2) + (\gamma^2 - b\gamma - wb) = 0
\label{quadeq}
\end{equation}
The larger root of this equation is:
\begin{equation}
f = \frac{(b - w - 2\gamma) + \sqrt{(b + w)^2 + 4w\gamma}}{2b}
\end{equation}
This expression is acceptable, i.e., positive, provided:
\begin{equation}
b( \gamma + w)/\gamma^2 > 1 \label{thres2}
\end{equation}
Thus Eq. (\ref{thres2}) serves as a threshold condition for the
existence of a non-trivial endemic level.

For $n=3$ it is not difficult to find a cubic equation for $f$
analogous to Eq. (\ref{quadeq}) (in this expression we set $\gamma=1$,
which sets the unit of time):
\begin{eqnarray}
f^3b^3-f^2(b^3-3wb^2-3b^2) & -& f(b(2b+3bw)-2bw^2-3b-3bw) \nonumber \\
-b(1+2w+2w^2)+1&=& g(f) =0 \label{cubiceq}
\end{eqnarray}
By inspection $g(1) >0$. If $g(0)= -b(1+2w+2w^2)+1 <0 $ then there is
at least one root in the interval $[0,1]$. (In fact, numerics shows
that there are also two negative roots). Thus if:
\begin{equation}
b(1+2w+2w^2)>1 \label{thres3}
\end{equation}
we have a non-trivial solution, and otherwise all the roots of
Eq. (\ref {cubiceq}) are all negative, and only the trivial solution
is acceptable.

In the general case we look for the stationary solution of
 Eq. (\ref{Timedep}) by setting $d\mathbf{P}/dt=0$. We get a
 homogeneous system of equations, from which the endemic level has to
 be established self-consistently. Recall that $\sum_kP_k=1$, so that
 we can reduce the number of equations by one by eliminating
 $P_0$. Denote the vector of length $n$, $\mathbf{p} =
 \left(P_{1},...,P_{n}\right)^T $. 
The resulting system of equations
 for the stationary state is inhomogeneous. It reads
\begin{eqnarray}
\mathbf{S}(f)\mathbf{p}&=&-nfb\mathbf{u} \nonumber \\ \mathbf{u} &=&
(1, 0, 0 , ...)^T \label{inhomogen}
\end{eqnarray}

The matrix $\mathbf{S}$ is given by:
\begin{equation}
\left(
\begin{array}{llllll}
m_{1,1}-nfb & \quad m_{1,2} -nfb & \quad -nfb &-nfb&... & -nfb \\
\quad m_{21} & \quad m_{22} & \quad m_{23} & 0 & ... & 0 \\ \quad
... &\quad ... & \quad ... & ... & ... & ... \\ \quad 0 & ... &\quad
m_{k,k-1} & m_{k,k} & m_{k,k+1} & 0... \\ \quad ... & ... & ... &
... & ... & ... \\ \quad 0 & 0 & ... & 0 & m_{n,n-1} & m_{nn}
\end{array}
\right)
\end{equation}

The matrix $\mathbf{S}(f)$ is linear in $f$:
\begin{equation}
\mathbf{S}(f)=\mathbf{W}+fb\mathbf{B}
\end{equation}
where both matrices, $\mathbf{W}$ and $\mathbf{B}$ are
$f$-independent.  The structure of $\mathbf{W}$ is:
\begin{equation}
\left(
\begin{array}{llll}
-\gamma -(n-1)w & \quad 2\gamma & \quad 0 & ... \\ 
(n-1)w & -2\gamma -2(n-2)w &\quad 3\gamma & ... \\ 
0 &\quad 2(n-2)w & -3\gamma -3(n-3)w &\quad 4\gamma \\ 
0 &\quad 0 &\quad 3(n-3)w & ... \\
0 &\quad  ... & \quad (n-1)w & -n\gamma 
\end{array}
\right)
\end{equation}
$\mathbf{W}$ depends only on the parameters $\gamma $ and $w$,
characterizing the infection within a household. Note that it is
nondegenerate: $\det \mathbf{W}=(-\gamma )^{n}n!$.  To see this, add
to each row the sum of all the rows below. The result is a lower
triangular matrix with diagonal elements in the $n$-th row equal to
$-n\gamma $. Moreover, $\mathbf{B}$ is a matrix with integer
components:
\begin{equation}
\mathbf{B=}\left(
\begin{array}{lllll}
-(2n-1) &\quad -n & -n & -n &\quad ... \\ 
(n-1) & \quad -(n-2) & \quad 0 & \quad 0 & \quad... \\ 
0 & \quad (n-2) & -(n-3) &\quad 0 &\quad ... \\ 
... & \quad ... & \quad ... &\quad  ... &\quad ... \\
 0 &\quad 0 & ... &\quad 1 &\quad 0
\end{array}
\right) .
\end{equation}

We give the explicit forms of the matrices $\mathbf{W}$ and
$\mathbf{B}$ for $n=2$ and $n=3$, setting $\gamma =1$: For $n=2$ one
has
\begin{equation}
\mathbf{W=}\left(
\begin{array}{ll}
-1-w & \quad 2 \\ \quad w & -2
\end{array}
\right) \mbox{ and }
\mathbf{B=}\left(
\begin{array}{ll}
-3 & -2 \\ \quad 1 & \quad0
\end{array}
\right) , \label{Two}
\end{equation}
For $n=3$ we get
\begin{equation}
\mathbf{W=}\left(
\begin{array}{lll}
-1-2w & 2 & 0 \\ 2w & -2-2w & 3 \\ 0 & 2w & -3
\end{array}
\right) \mbox{ and }\mathbf{B=}\left(
\begin{array}{lll}
-5 & -3 & -3 \\ 2 & -1 & 0 \\ 0 & 1 & 0
\end{array}
\right) .  \label{Three}
\end{equation}

From the remarks above, it is clear that there is a non-zero solution,
$\mathbf{p}$, to Eq. (\ref{inhomogen}) for any $f>0$, so that
$\mathbf{S}^{-1} $ exists. Thus:
\begin{equation}
\mathbf{p}=-nfb\mathbf{S}^{-1}\mathbf{u} \label{Ex}
\end{equation}

From this equation a closed form for the endemic level $f$ follows.
Multiplying both sides of Eq.(\ref{Ex}) by the vector $\mathbf{n_1}
=(1,2,...,n)$ and noting that $f=n^{-1}\mathbf{n_1p}$ we get:
\begin{equation}
f=-fb\mathbf{n_1}\mathbf{S} ^{-1}\mathbf{u}
\end{equation}
which is a closed algebraic equation for $f$ which always has the
trivial solution $f=0$.  Factoring out this out, we get an equation
for the nontrivial solution:
\begin{equation}
-b\mathbf{n_1}\mathbf{S} ^{-1}\mathbf{u}=1.  \label{Ef}
\end{equation}

As in Eqs. (\ref{thres2}, \ref{thres3}) there is a condition on the
parameters for the solution to be positive. For given $w$ we define a
value $b_c$ such that for $b>b_c$ we have such an $f$. This can be
found from Eq.(\ref{Ef}) by noting that at $b_{c}$ the second
solution crosses $f=0$. At that point, $\mathbf{S=W}$, and we get:
\begin{equation}
b_{c}=-1/\left[ \mathbf{n_1W}^{-1}\mathbf{u}\right] \label{bc }.
\end{equation}

To get an explicit expression for $b_{c}$ consider the vector $
\mathbf{v=W}^{-1}\mathbf{u}$, whose elements are $\mathbf{v}_{k}
\mathbf{=}\left[ \mathbf{W}^{-1}\right] _{1,k}$. As we will show,
\begin{equation}
\mathbf{v}_{k}=-\frac{(n-1)!}{k(n-k)!}w^{k-1}, \label{Matrelem}
\end{equation}
so that the threshold is given by
\begin{equation}
b_c^{-1}=-\mathbf{n_1v=}\sum_{k=1}^{n}\frac{(n-1)!}{(n-k)!}w^{k-1}.
\label{thresh}
\end{equation}

To prove Eq.(\ref{Matrelem}) we note that the solution to
$\mathbf{Wv=u}$ is unique so that we need only insert
Eq. (\ref{Matrelem}) into $\mathbf{Wv=u}$ . The first element is
\begin{equation}
-(-1-(n-1)w)-(n-1)w=1,
\end{equation}
while the $k$-th element ($k>1$) is:
\begin{eqnarray}
&&(k-1)(n+1-k)w\cdot \frac{(n-1)!}{(n+1-k)!(k-1)}w^{k-2} \nonumber \\
&&+(-k-k(n-k)w)\frac{(n-1)!}{(n-k)!k}w^{k-1}+ \nonumber \\
&&+(k+1)\frac{(n-1)!}{(n-k-1)!(k+1)}w^{k}= \\ &=&(n-1)!\left\{ \left[
\frac{1}{(n-k)!}-\frac{1}{(n-k)!}\right] w^{k-1}\right. + \nonumber \\
&&+\left. \left[ \frac{1}{(n-k-1)!}-\frac{1}{(n-k-1)!}\right]
w^{k}\right\} \nonumber \\ &=&0.  \nonumber
\end{eqnarray}

Once we obtain $\mathbf{p}$ we can find other moments of the
distribution by multiplying Eq.(\ref{Ex}) by $\mathbf{n_q} =
(1,2^q,3^q, ...,n^q)$. We present numerical results on $\mathbf{n_2
p}$ below.

We now discuss the behavior of the solution in the vicinity of the
transition point. Eq.(\ref{Ex}) can be rewritten:
\begin{equation}
\mathbf{p=}-nfb\left( \mathbf{I}+fb\mathbf{W}^{-1}\mathbf{B}\right)
^{-1} \mathbf{W}^{-1}\mathbf{u,}
\end{equation}
Then we formally expand in a geometric series:
\begin{equation}
\mathbf{p=}-nfb\left(
\mathbf{I}-fb\mathbf{W}^{-1}\mathbf{B+(}fb\mathbf{)}^{2}\left(
\mathbf{W}^{-1}\mathbf{B}\right) ^{2}-...\right) \mathbf{W}^{-1}
\mathbf{u.}  \label{iterat}
\end{equation}
Multiplying both sides of Eq.(\ref{iterat}) by $n^{-1}\mathbf{n_1}$ we
get the following equation for $f$:
\begin{equation}
f=-fb\left(
a_{0}-fba_{1}+\mathbf{(}fb\mathbf{)}^{2}a_{2}+...+(-1)^{m}
\mathbf{(}fb\mathbf{)}^{m}a_{m}+...\right) \label{selfcons}
\end{equation}
where
\begin{equation}
a_{m}=\mathbf{n_1}\left( \mathbf{W}^{-1}\mathbf{B}\right)
^{m}\mathbf{W}^{-1}\mathbf{u} \label{coeff}
\end{equation}
from which $f$ can be obtained with any necessary accuracy. Note that
according to Eq.(\ref{bc }) the coefficient
\begin{equation}
a_{0}=\mathbf{n_1W}^{-1}\mathbf{u}=-1/b_{c}.  \label{a0}
\end{equation}
Eq.(\ref{selfcons}) is especially useful in the vicinity of
transition, where $f$ is small. Both the point of transition and the
endemic level near the threshold can be determined.

Eq.(\ref{selfcons}) always possesses a trivial solution $f=0$. An
equation for a nontrivial one then reads:
\begin{equation}
G(f)=fb^{2}a_{1}-f^{2}b^{3}a_{2}+...+(-1)^{m+1}f^{m}b^{m+1}a_{m}+...=1+ba_{0}.
\end{equation}
The formal solution of this equation involves the inverse of $G$:
\begin{equation}
f=G^{-1}(1+ba_{0})=G^{-1}(1-b/b_{c}).
\end{equation}
where Eq (\ref{a0}) has been used. Using the relationship between the
Taylor series of a function and its inverse we get, in the vicinity of
$b_{c}$:
\begin{equation}
f=-\beta _{1}\tau +\beta _{2}\tau ^{2}-...
\end{equation}
where $\tau =(b-b_{c})/b_{c}$ and the coefficients are:
\begin{eqnarray}
\beta _{1} &=&\frac{1}{b^{2}a_{1}(w)}, \nonumber \\ \beta _{2}
&=&-\frac{a_{2}(w)}{b^{3}a_{1}^{3}(w)}, \\ \beta _{3}
&=&\frac{1}{b^{4}a_{1}^{5}(w)}\left( 2a_{2}(w)-a_{1}(w)a_{3}(w)\right)
, \nonumber \\ &&...  \nonumber
\end{eqnarray}
In the vicinity of the transition one can take $b=b_{c}(w)$; the
 behavior of $f$ is dominated by the linear term:
\begin{equation}
f=-\beta _{1}\tau =-\frac{1}{b^{2}a_{1}(w)}\tau
\end{equation}
Explicit calculations show that $a_{1}<0$.

\subsection{Stability of the disease-free equilibrium}

Restoring the time-dependence in Eq.(\ref{inhomogen}) we obtain an
equation for the time evolution of the reduced probability vector
$\mathbf{p}$:
\begin{equation}
\frac{d}{dt}\mathbf{p}=(\mathbf{W}+fb\mathbf{B})\mathbf{p}+nfb\mathbf{u.}
\label{temporal}
\end{equation}
Since $\mathbf{p}$ depends on $f$, Eq.(\ref{temporal}) is a nonlinear
equation describing the evolution of $\mathbf{p}$.  In the previous
section we have shown that this equation has two equilibria provided
$b > b_c$, where $b_c$ is defined in Eq. (\ref{thresh}). We will now
show that under the same conditions the trivial solution, $f=0,
\mathbf{p=0}$ is unstable.

In the vicinity of the trivial equilibrium, we can omit the term
$fb\mathbf{B}$ in brackets in Eq. (\ref{temporal}). The linearized
equation reads
\begin{equation}
\frac{d}{dt}\mathbf{p}=\mathbf{Wp}+b(\mathbf{n_{1}p})\mathbf{u} \equiv
\mathbf{Tp}
\label{linearized}
\end{equation}
where
\begin{equation}
\mathbf{T}=\mathbf{W}+b\mathbf{N,} \label{MatrT}
\end{equation}
The matrix $\mathbf{N}$ has $\mathbf{n_1}$ as its first row and all
other elements zero.

The determinant of $\mathbf{T}$ is linear in $b$. To see
this, note that the explicit form of $\mathbf{T}$ is:
\begin{equation}
\left(
\begin{array}{llll}
-\gamma -(n-1)w+b &\quad 2\gamma +2b & \quad3b & ... \\
 (n-1)w & -2\gamma -2(n-2)w &\quad  3\gamma & ... \\
 0 & 2(n-2)w & -3\gamma -3(n-3)w &... \\
... & ... & ... & ... \\ 0 & ... & (n-1)w & -n\gamma
\end{array}
\right)
\end{equation}
If we expand the determinant about the first row, each term contains
$b$ only once. It can thus be written $\det \mathbf{T}=\det
\mathbf{W}+cb$, with $c$ a constant.  However, the determinant of
$\mathbf{T}$ is the product of its eigenvalues.  Since for $b=0$
(separate households) the trivial solution is stable, all of the
eigenvalues of $\mathbf{W}$ are negative; $\det \mathbf{W}$ is
nonzero, and its sign is positive for $n$ even and negative for $n$
odd. From the linearity in $b$ we see that $\det \mathbf{T}$ changes
sign at $b=-c^{-1}\det \mathbf{W}$ which corresponds to the appearance
of a positive eigenvalue.

To find the $b$ for which $\det \mathbf{T}=0$ write Eq.(\ref{MatrT})
as:
\begin{equation}
\mathbf{TW}^{-1} = \mathbf{I} + b \mathbf{NW}^{-1}.
\label{tw}
\end{equation}
The determinant of the matrix in the l.h.s. of this expression is
\begin{equation}
\det \mathbf{T/}\det \mathbf{W=}\left[ \mathbf{(-\gamma
)}^{-n}/n!\right] \det \mathbf{T} \nonumber
\end{equation} 
and vanishes when $\det \mathbf{T}=0$.  But, since $\mathbf{N}$ has
non-zero elements only in the first row, the matrix on the r.h.s. of
Eq. (\ref{tw}) is:
\begin{equation}
\left(
\begin{array}{lllll}
1+bq_{1} &\quad  bq_{2} &\quad  bq_{3} &\quad ... &\quad bq_{n} \\
0 & \quad 1 &\quad  0 &\quad ... &\quad ... \\
0 & \quad 0 & \quad 1 &\quad  0 & \quad ... \\
 ... &\quad ... &\quad ... &\quad ... & \quad... \\
 0 &\quad 0 & \quad... &\quad 0 &\quad 1
\end{array}
\right) ,
\end{equation}
where $q_{i}$ is the $i$-th element of the vector
$\mathbf{n_1W}^{-1}$.  By inspection, the determinant of this matrix
is $1+q_{1}\equiv 1+b\mathbf{n_1W}^{-1} \mathbf{u}$. It vanishes at
$b=-1/(\mathbf{n_1W}^{-1}\mathbf{u})$, i.e.  exactly at $b=b_{c}$, see
Eq.(\ref{bc }). Therefore, for $b>b_{c}$ a positive eigenvalue
appears, and the trivial solution loses its linear stability.

\subsection{Numerical results}
\begin{figure}
\centerline{\includegraphics[width=5in]{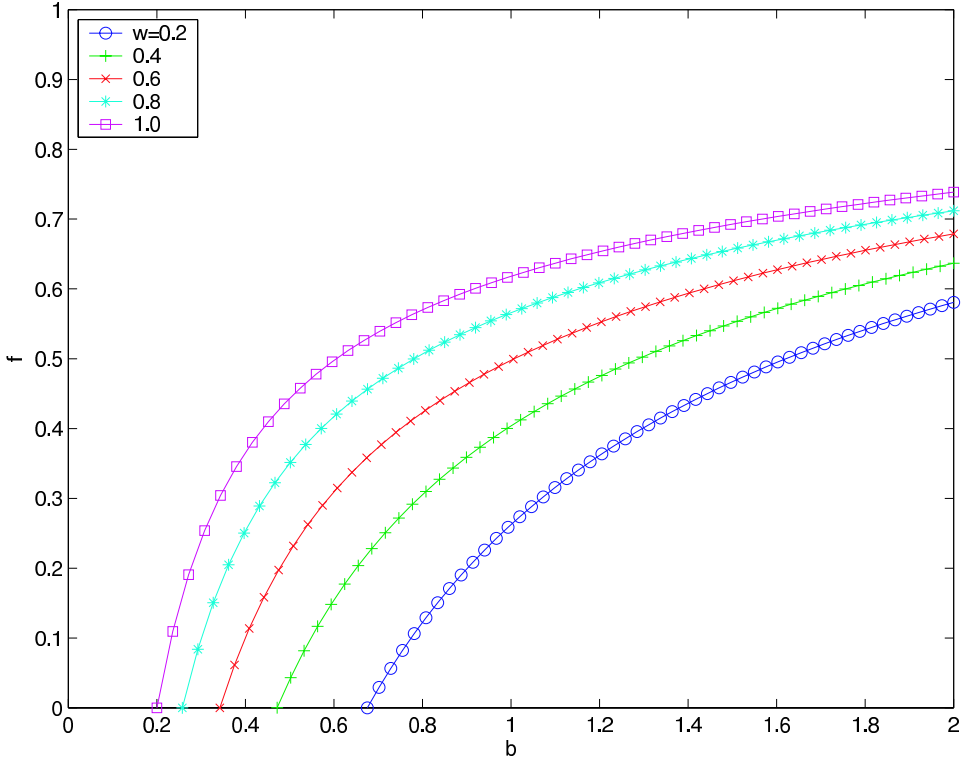}}
\caption{The endemic level as a function of $b$ for $n=3$ and various
$w$, $\gamma=1$}
\label{fig1}
\end{figure}

\begin{figure}
\centerline{\includegraphics*[width=5in]{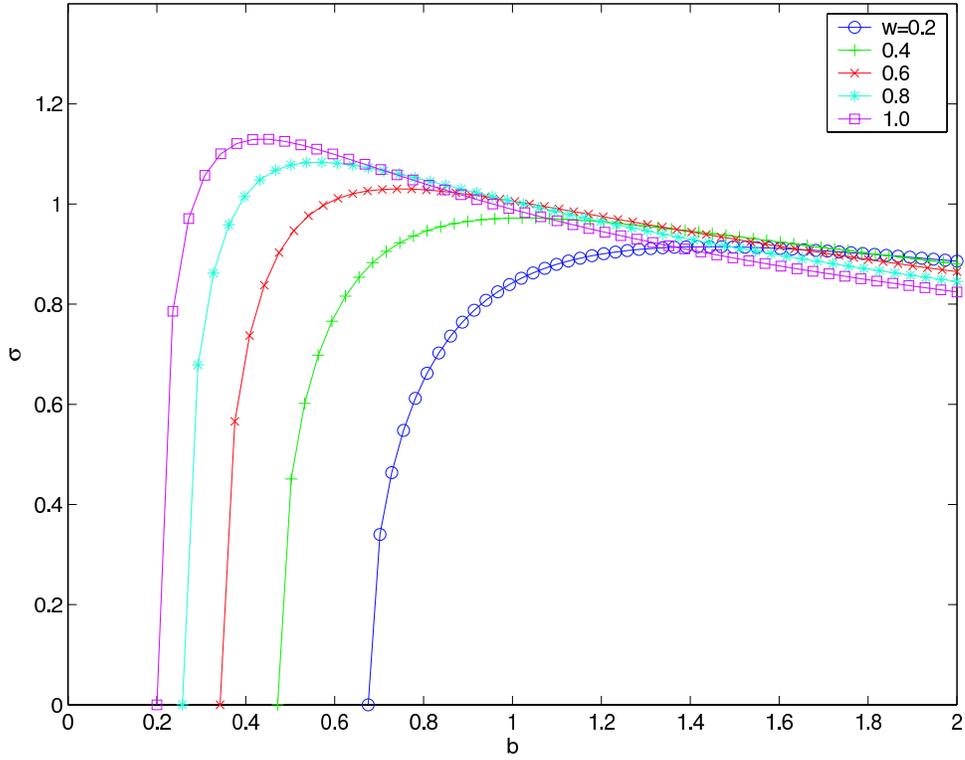}}
\caption{The standard deviation of the number of infected in a
household, as a function of $b$ for $n=3$ for several $w$'s,
$\gamma=1$}
\label{fig2}
\end{figure}

\begin{figure}
\centerline{\includegraphics[width=5in]{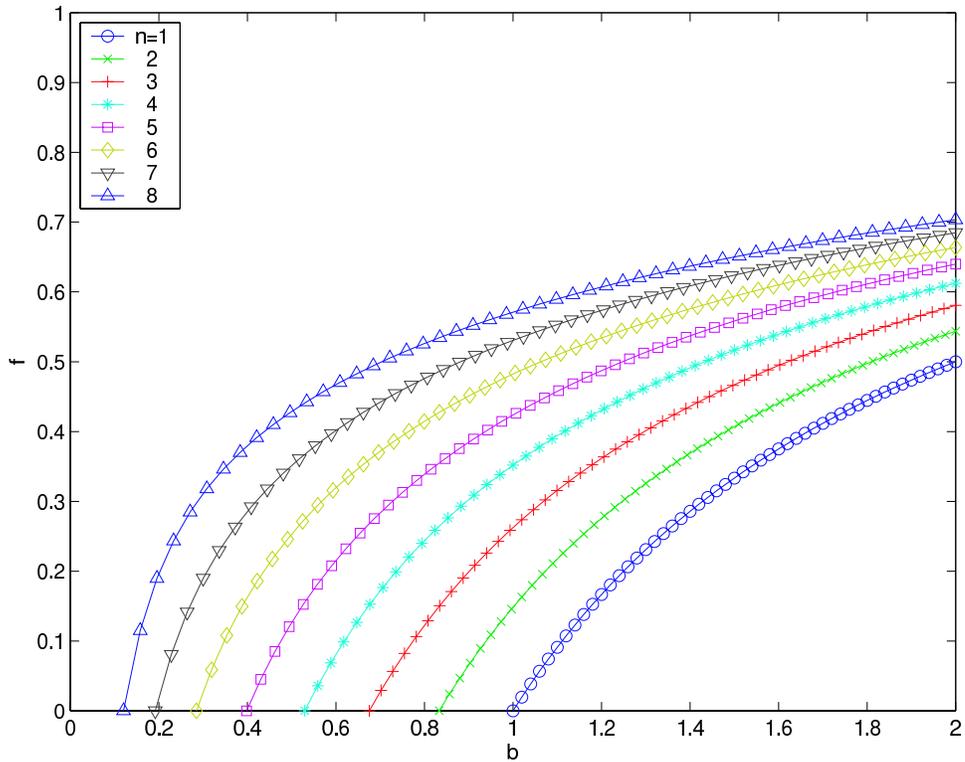}}
\caption{The endemic level as a function of $b$ for $w=0.2$ and
$n=1,2,..8$, $\gamma=1$}
\label{fig3}
\end{figure}

\begin{figure}
\centerline{\includegraphics*[width=5in]{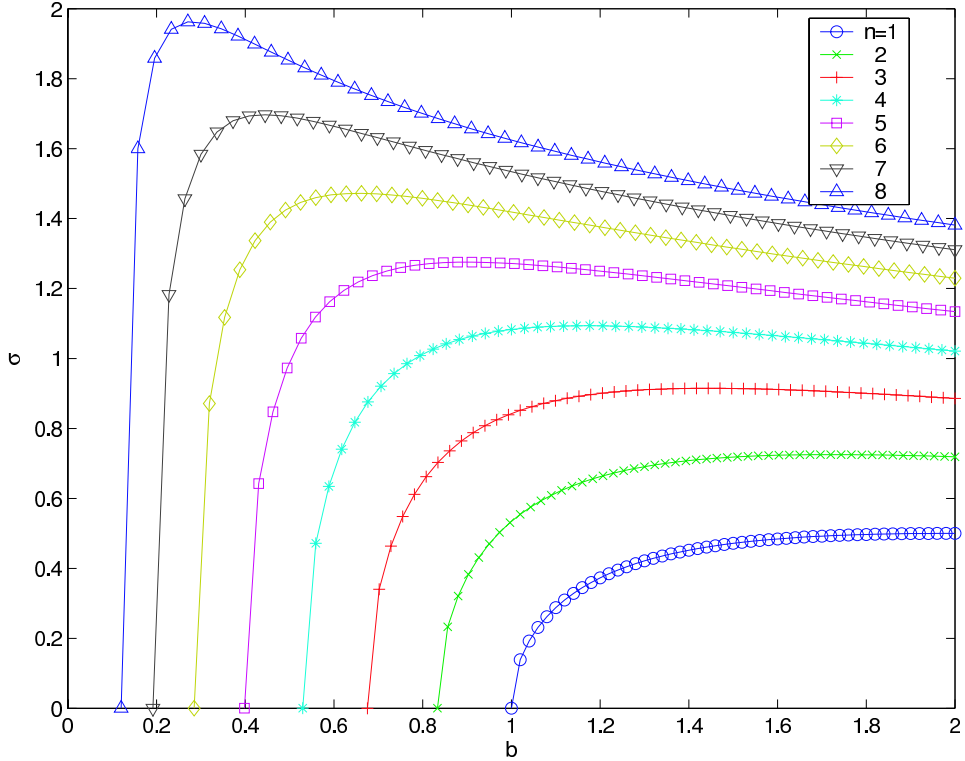}}
\caption{The standard deviation of the number of infected in a
household, as a function of $b$ for $w=0.2$ and $n=1,2, ...,8$,
$\gamma=1$}
\label{fig4}
\end{figure}

It is quite simple to work out the dependence of $f$ on the parameters
$w, b, n$. We wrote a Matlab program which solved Eq. (\ref{Ef}).  In
Figure (\ref{fig1}) we show $f$ as a function of $b$ for several $w$'s
for $n=3$.  In this and the following figures we have taken
$\gamma=1$. For other values of $\gamma$ one can use the obvious
scaling relation $(w,b,\gamma) \to (w/\gamma, b/\gamma, 1)$.

Recall that $f=\mathbf{n_1p}$ can be interpretated as $<i>/n$ where
$i$ is the number infected in a household, and $<>$ is the average
over the distribution $\mathbf{P}$.  If we calculate $\mathbf{n_2p}$
we get a measure of the fluctuation of $i$ in a household. In Figure
(\ref{fig2}) we show the standard deviation, $\sigma = [<i^2> -
<i>^2]^{1/2} = [\mathbf{n_2p} - (nf)^2]^{1/2}$ as a function of $b$
for several $w$'s, for $n=3$.  In a similar way, in Figures
(\ref{fig3}, \ref{fig4}) we show $f$ and $\sigma$ for $w=0.2$ and
$n=1,2,..., 8$.

\section{Simulations: finite systems and finite range}
The considerations in the previous sections apply in the limit $m \to
\infty$. For any finite $m$ the only steady-state solution possible is
the trivial one, $f=0$; with probability unity the infection will
become extinct. However, the time to extinction may be very long,
$O(e^{m})$ if the system is above $b_c$. See, for example,
\cite{Weiss,Nasell,Swedes} for the case $n=1$. For times short
compared to the extinction time the system will be in a
quasi-stationary state which resembles the infinite system.

\begin{figure}
\centerline{\includegraphics[width=3in]{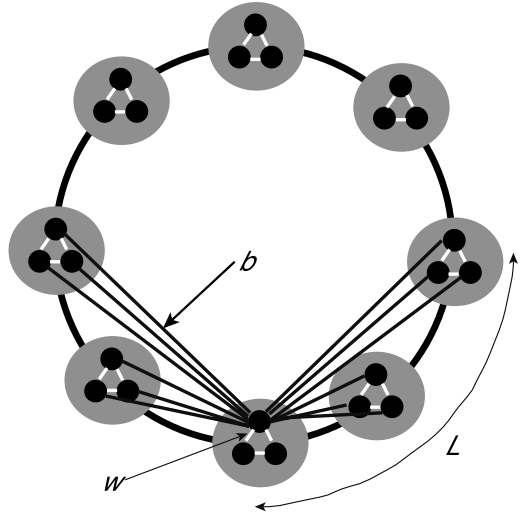}}
\caption{The coupling scheme used in the simulations. This is the case
$n=3$, $m=8$, $L=2$. The white bonds represent contacts within a
household. All of the inter-household contacts for one agent are shown
in black}
\label{sketch}
\end{figure}

We have done preliminary investigations of this and related behavior
using computer simulations. We represent the population as follows:
imagine a population of $m$ households each containing $n$ agents
arranged on a ring. All of the members of the same household can
infect one another with probability/unit time $w$. The communication
with other households is variable range: we allow each household to
infect all of the other households in a range $L$ on either side. See
Figure (\ref{sketch}). That is, the number of outside contacts is
$q=2Ln$, and the infection probability is $b/q$. The equivalent of the
random mixing case of the previous sections is $2L=m-1$; in this case
all agents are coupled to each other. The case $n=1, L=1$ is the 
well-known contact process
\cite{contact}.

\begin{figure}
\centerline{\includegraphics[width=5in]{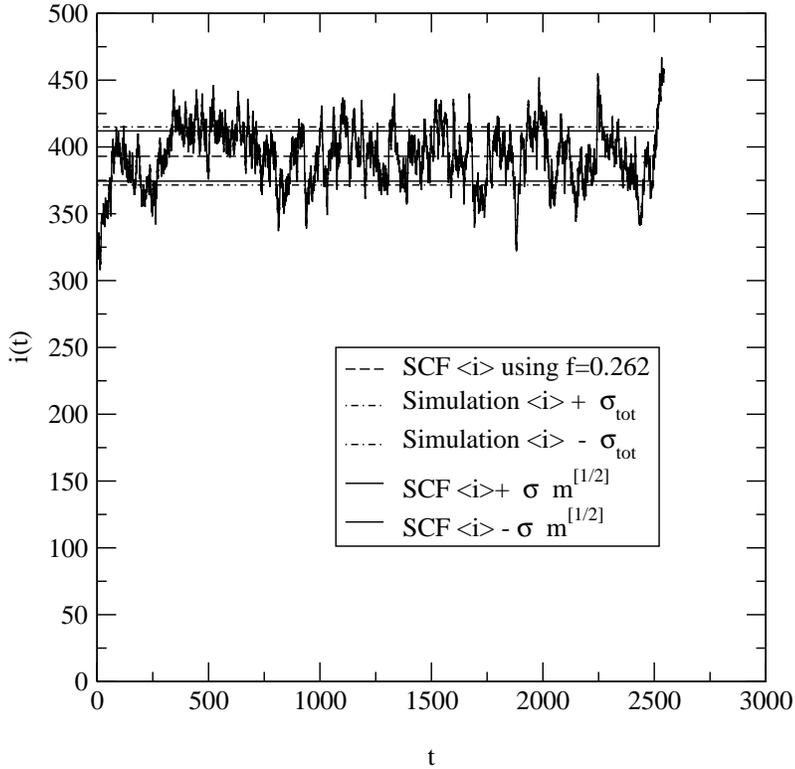}}
\caption{Simulation results for the number of infected as a function
of time compared to the SCF results. $n=3$,
$m=501$, $L=150$, $w/\gamma=0.2, b/\gamma=1$. The agreement for the
average is essentially perfect, but for the fluctuations (see text)
the SCF method is an underestimate.}
\label{sim}
\end{figure}

Our computer algorithm is quite simple: we suppose that $\gamma, w, b$
are all less than unity.  We start by marking a certain fraction of
agents as infected, which establishes $I(0)$ where $I$ is the total
number infected in the population. For each computer time step we
choose one agent, in household $k$, at random among the infected, and
find all the other agents that are coupled to the one in
question. These will be in households with index $k-L, k-L+1,
...,k+L$, where the indicies are to be interpreted with periodic
boundary conditions. With probability $w$ or $b/q$ (depending whether
the other agent is in $k$ or outside) the other agent is
infected. Then another agent is chosen at random among the infected,
and it is allowed to recover with probability $\gamma$.  Each computer
time step corresponds to an increment of `clock time', $dt= 1/I$. A
typical result for $I(t)$ is shown in Figure(\ref{sim}). It is worth
noting that this algorithm is quite fast -- it is not difficult to
treat as many as 3000 agents on a simple workstation.

In this section we will give a few results of the simulations. We have
found that the SCF is an excellent representation of the
behavior in large systems with long-range interactions, but that for
smaller systems and shorter ranges the approximation becomes poorer,
but still surprisingly good. A
systematic investigation of these effects will be reserved for a
future publication: here we give a few preliminary results.

\subsection{Finite size effects}
For large enough $m$ the simulation results agree very well with the
SCF results. In Figure(\ref{sim}) the average of the time
series agrees almost perfectly with the results from the previous
sections: the numerical average of the simulation data differs from the
SCF by less than the width of the line.

In addition we can try to understand the fluctuations of the time
series by the following reasoning: if we imagine that the different
households fluctuate more or less independently (this is the
assumption under which the SCF method is
valid), then the standard deviation of the total should be
$\sigma_{tot}=\sigma \sqrt{m}$. As we see in the figure, this is a
good approximation. However, we have observed that there is a large
finite-size effect for this statistic. For $m=301$ the simulation
result for $\sigma_{tot}$ is almost twice the prediction. 

Another way to look at the standard deviation is to simply record the time 
series of the number infected in a given household, and directly find $\sigma$.
We did this by averaging over 10 randomly chosen households in the
system. We find very good agreement with the analytic treatment: for
example, for $w=0.2, b=1., n=3, m=501$ we find $\sigma=0.82$. The analytic
result is $0.84$, well within our estimated error.

As we reduce $m$ we find a surprising result: within the 
numerical accuracy of our simulation  we can find 
no change in $f$. Rather, at a certain small size, $m=50$ for the parameters
in the last paragraph, the infection dies before we can collect adequate 
statistics. 

\subsection{Finite range effects}
For the case of small $L$ the SCF ceases to
be valid. In magnetic systems, locality of the interaction is well
known to increase fluctuations and to change the threshold. We find
related effects here, but  the range needs to be very small indeed
for the SCF result to be very inaccurate.

\begin{figure}
\begin{center}\includegraphics[width=6in]{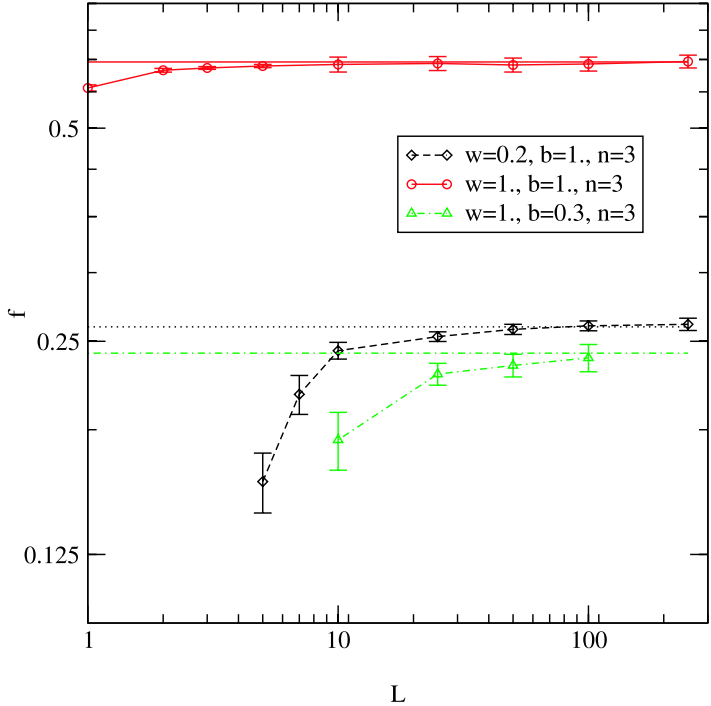}
\end{center}
\caption{The endemic level, $f$, as a function of the range
of the infection contacts between  households. The horizontal
lines are the SCF results for large $L$. In all cases $n=3, m=501$
so that the maximum possible  $L=250$}
\label{fofL}
\end{figure}

In Figure (\ref{fofL}) we show simulation results for $f(L)$, the
endemic level as a function of the range of the interaction 
between households, $L$. We show  three different 
sets of parameters for $n=3$ and compared with the SCF ($L \to \infty$).
The dependence is remarkable: for the parameter values ($w=1., b=1.$), 
finite range has hardly any effect. The other two sets do show a decrease
of $f$, and an apparent shift in the threshold. For $w=0.2$ 
$b_c=0.67$ in the SCF, well below $b=1.$ However, for $L < 5$, the system is 
below threshold, and the epidemic dies quickly. A similar effect occurs 
for $w=1., b=0.3$, which is below $b_c$ for $L<10$. (For large $L$, 
$b_c =0.2$ for these parameters.) Another way to put this is that there
is a characteristic length $L_c \approx 10$ for these two sets
of parameters.

These results are reminiscent of effects that occur in phase
transitions in statistical physics. In fact, the contact process is
fruitfully viewed as an example of a 
continuous non-equilibrium phase transition \cite{Dickmann}.
Near the threshold (i.e. the critical point) of a continuous phase 
transition, the system develops a diverging correlation length. For
total system sizes greater than this length finite size effects are
small. Thus we might guess that the correlation lengths for two of
the sets of parameters could be of order
$L_c$. We have not investigated this matter in detail, but it would
repay consideration.

\section{Summary}
In this paper we have given a number of explicit results for the SIS household
model. Our point of view was to emphasize the
SCF method as a guide to solution and insight.
The numerical results give rise to a number of questions that are probably not 
accessible to rigorous proof, but for which numerical methods could be pursued. 
For example, we have speculated that there are critical fluctuations 
\cite{Huang,Dickmann} in this system. That would mean that near threshold
there would be large correlations in adjacent households for a finite range
system, a result that could be of considerable interest.

There is also a practical aspect to this method.  SCF theory is quite flexible, and allows considerable complication to  be added to the model without substantially increasing the 
complexity of the solution. Suppose our household contains two types of members, for example, children who go to school and contract infections, and adults who do not. This would require a few changes: for a 3 person household with two adults and a child, the analog of the vector 
$\mathbf{P}$ would be of length 6 instead of 4, 
and the complete solution would be no worse than the numerical inversion of a $5\times 5$ matrix. 

\ack 
We would like to thank Frank Ball, Mark Newman, Carl Simon, and Jim Koopman for 
helpful discussions. GG was supported by the REU summer program of the
Physics Department of the University of Michigan, and IMS would like to 
thank the Michigan Complex Systems program for support.

\end{document}